\documentclass[twocolumn,showpacs]{revtex4}
\usepackage{graphicx}
\usepackage{dcolumn}

\begin{document}


\title{Shape transition and oblate-prolate coexistence in $N=Z$ $fpg$-shell nuclei}

\author{K. Kaneko$^{1}$, M. Hasegawa$^{2}$, and T. Mizusaki$^{3}$}
\affiliation{
$^{1}$Department of Physics, Kyushu Sangyo University, Fukuoka 813-8503, Japan \\
$^{2}$Laboratory of Physics, Fukuoka Dental College, Fukuoka 814-0193, Japan \\
$^{3}$Institute of Natural Sciences, Senshu University, Kawasaki, Kanagawa, 
214-8580, Japan
}

\date{\today}

\begin{abstract}
Nuclear shape transition and oblate-prolate coexistence in $N=Z$ nuclei are investigated 
within the configuration space ($2p_{3/2}$, $1f_{5/2}$, $2p_{1/2}$, and  $1g_{9/2}$). 
We perform shell model calculations for $^{60}$Zn, $^{64}$Ge, and $^{68}$Se and 
constrained Hartree-Fock (CHF) calculations for $^{60}$Zn, $^{64}$Ge, $^{68}$Se, and $^{72}$Kr, 
employing an effective pairing plus quadrupole residual interaction with monopole interactions. 
The shell model calculations reproduce well the experimental energy levels of these 
nuclei. From the analysis of potential energy surface in the CHF calculations, 
we found shape transition from prolate to oblate deformation in these $N=Z$ nuclei 
and oblate-prolate coexistence at $^{68}$Se. The ground state of $^{68}$Se has oblate shape, 
while the shape of $^{60}$Zn and $^{64}$Ge are prolate. It is shown that the isovector 
matrix elements between $f_{5/2}$ and $p_{1/2}$ orbits cause the oblate 
deformation for $^{68}$Se, and four-particle four-hole ($4p-4h$) excitations are important 
for the oblate configuration. 

\end{abstract}


\maketitle

The proton-rich nuclei with masses between 56 and 80 are well known to 
change rapidly their collective properties with proton and neutron numbers. 
The relevant Nilsson diagram \cite{Nazare} shows a large shell gap for $N$, $Z=$34 at prolate 
and oblate deformations, for $N$, $Z=$36 at large oblate deformation, and for $N$, $Z=$38 
at large prolate deformation. Particular interests are in shape transition and 
oblate-prolate coexistence of heavier $N=Z$ nuclei. 
These nuclei lie in the transitional region from the spherical shape 
(e.g. $^{56}$Ni \cite{Rudolph1}) with coexisting prolate shape \cite{Mizusaki} 
to the prolate deformation (e.g. $^{80}$Zr \cite{Lister}). 
The nucleus $^{64}$Ge \cite{Ennis} is known to be an $N=Z$ 
proton-rich unstable nucleus manifesting 
a $\gamma$-soft structure from the theoretical calculations based on the mean-field approximation. 
The early studies of $^{69}$Se \cite{Wiosna} have found 
indications of oblate shapes. Long standing predictions of a stable oblate deformation were 
recently confirmed by the observation of an oblate ground state band in $^{68}$Se 
\cite{Fischer}. Determinations of shape were inferred indirectly from the study of rotational 
bands, while direct quadrupole measurements are difficult for short-lived states. 
In their analyses, it was suggested that the oblate 
configuration coexists with a prolate rotational band which is so-called 
oblate-prolate coexistence. 
Previous work \cite{Angelis} for $^{72}$Kr suggests that the ground state has an oblate shape 
and there is an oblate $\rightarrow$ prolate shape transition at low spins. 
In the total Routhian surface (TRS) calculations \cite{Stefanova}, the oblate minimum can be 
understood in terms of the large gap for oblate deformation at $N=Z=$36. 
Projected shell model \cite{Palit}, a deformed selfconsistent Hartree-Fock (HF)+RPA approach 
\cite{Sarrig}, and Skyrme Hartree-Fock-Bogolyubov (HFB) calculations \cite{Yamagami} for 
proton-rich nuclei in the $A=$70-80 mass region were performed. 
Recent investigations for $^{68}$Se have been carried out by the projected shell 
model \cite{Sun}, by the excited VAMPIR method \cite{Petrovici}, and the self-consistent 
collective coordinate method \cite{kobayasi}. 

The spherical shell model approaches could be more appropriate for describing various 
aspect of nuclear structure. However, the shell model calculations 
by diagonalization in the $1f_{7/2}, 2p_{3/2}, 1f_{5/2}, 2p_{1/2}$, 
and $1g_{9/2}$ for nuclei with $N, Z=30-36$ are hopeless at present because 
of huge dimension of configuration space. 
So we need to restrict the model space to the $2p_{3/2}, 1f_{5/2}, 2p_{1/2}$, and $1g_{9/2}$ 
orbits (henceforth called $fpg$-shell). 
The dimension of the configuration space is still huge. For instance, maximum dimension 
for $^{68}$Se is 0.165 billion. 
Recently, an extended pairing plus quadrupole-quadrupole ($P+QQ$) force has been applied 
to the $fpg$-shell nuclei, and shown to be useful \cite{Kaneko,Hasegawa}. 
This interaction works remarkably well. 
In this paper, we study the shape transition and oblate-prolate coexistence in the 
$N=Z$ $fpg$-shell nuclei using a large scale shell model 
and the constrained Hartree-Fock (CHF) calculations. 
The shell model calculations with $\sim 10^8$ dimension can be 
carried out by recently developed shell model code \cite{Mizusaki1}. 
The observation associated with oblate shape is rare compared with 
the prolate shape well established experimentally. 
The reason for the suppression of oblate deformation lies in 
the higher order effects both in liquid drop terms and residual interactions 
which make most nuclei favor prolate shapes. 
Very strong oblate-driving effects seem to be necessary for overcoming this prolate 
tendency. As a candidate for the oblate-driving force, we present the isovector ($T=1$) 
interaction with matrix element between $f_{5/2}$ and $p_{1/2}$ orbits. 

We start from the extended $P+QQ$ model \cite{Kaneko,Hasegawa} with the monopole 
interactions $V_{\rm m}$
\begin{eqnarray}
H & = & H_{0} + H_{P_{0}} + H_{QQ} + H_{OO} + V_{\rm m} \nonumber \\
 & = & \sum_{\alpha}\varepsilon_{a}c_{\alpha}^{\dag}c_{\alpha}-\frac{1}{2}g_{0}
 \sum_{\kappa}{P}^{\dag}_{001\kappa}{P}_{001\kappa} \nonumber \\
  & {} &  -\frac{1}{2}\chi_{2}\sum_{M}:{Q}^{\dag}_{2M}{Q}_{2M}: 
          -\frac{1}{2}\chi_{3}\sum_{M}:{O}^{\dag}_{3M}{O}_{3M}: \nonumber \\
  & {} &  - k^{0}\sum_{JM,ab}{A}^{\dag}_{JM00}(ab){A}_{JM00}(ab) + V_{\rm m}, 
  \\
  & {} & {P}^{\dag}_{001\kappa}=\sum_{a}\sqrt{j_{a}+1/2}{A}^{\dag}_{001\kappa}(aa), \\
  & {} &   {A}^{\dag}_{JMT\kappa}(ab)=[c_{a}^{\dag}c_{b}^{\dag}]_{JMT\kappa}/\sqrt{2}, 
\end{eqnarray}
where $\varepsilon_{a}$ is a single-particle energy, 
${P}_{001\kappa}$ is the $T=1,J=0$ pair operator, and 
${Q}_{2M}$ (${O}_{3M}$) is the isoscalar quadrupole (octupole) operator. 
Each term includes {\it p-n} components which play important 
roles in the $N=Z$ nuclei, due to an isospin-invariance. 

The shell model calculations \cite{Kaneko,Hasegawa} have been performed in a shell model space, 
which assumes a closed $^{56}$Ni core. 
The neutron single-particle energies of $2p_{3/2}$, $1f_{5/2}$, $2p_{1/2}$, and  $1g_{9/2}$ 
in this $fpg$-shell region can be read from the low-lying states of $^{57}$Ni. 
We used the measured values $\varepsilon_{p3/2}=0.0$, $\varepsilon_{f5/2}=0.77$,
$\varepsilon_{p1/2}=1.11$, and $\varepsilon_{g9/2}=3.70$
in MeV, in our previous paper \cite{Kaneko}. 
However, we have recently obtained a better single particle energy
$\varepsilon_{g9/2}=2.50$ MeV which fits odd-mass Ge isotopes in Refs.
\cite{Hasegawa1,Hasegawa2}. This value 
 is consistent with the report in \cite{Aberg}. 
Since the above Hamiltonian is assumed to be an 
isospin-invariant, the proton single-particle energies are taken
as the same values as the neutron single-particle energies. 
We searched force strengths of the extended $P+QQ$ interaction
using $\varepsilon_{g9/2}=2.50$ MeV. 
The obtained values are as follows: 
\begin{eqnarray}
g_{0}=0.270(64/A), \hspace{0.5cm} \chi_{2}=0.250(64/A)^{5/3}/b^{4}, \nonumber \\ 
\chi_{3}=0.05(64/A)^{2}/b^{6}, \hspace{0.5cm} k^{0}=1.44(64/A),
\end{eqnarray}
where $g_{0}$, $\chi_{2}$, $\chi_{3}$, and $k^{0}$ are the $J=0$ pairing, 
the $QQ$, the octupole, and $J$-independent isoscalar force strengths, 
respectively. Here $b$ is the harmonic-oscillator range parameter. 
The monopole shifts are 
\begin{eqnarray}
 & {} &  V_{m}(p_{3/2},f_{5/2};T=1)=V_{m}(p_{3/2},p_{1/2};T=1)=-0.3, \nonumber \\
 & {} &  V_{m}(f_{5/2},p_{1/2};T=1)=-0.4, \nonumber \\ 
 & {} &  V_{m}(g_{9/2},g_{9/2};T=1)=-0.2, \nonumber \\
 & {} &  V_{m}(g_{9/2},g_{9/2};T=0)=-0.1, \quad \mbox{ in MeV}. 
\end{eqnarray}
All these force strengths have been phenomenologically adjusted so as to fit 
many energy levels including high-spin levels for 
$^{60,62,64,66,68}$Zn,  $^{64,66,68,70}$Ge
and also $^{65,67}$Ge \cite{Hasegawa1,Hasegawa2}. 
The force strengths determined in this way describe quite well
various properties of these nuclei. 
In particular, the monopole shifts are important for describing
 precise positions of the high-spin ($J \ge 8$) states for $^{66,68}$Ge.

In Fig. 1, energy spectra calculated with the above force strengths 
are compared with experimental data in the $N=Z$ $fpg$-shell nuclei, $^{60}$Zn, $^{64}$Ge, 
and $^{68}$Se. 
In addition to the ground state band,  the second positive-parity 
band beginning from $J^{\pi}=2_{2}^{+}$ state and negative-parity band
beginning from $J^{\pi}=3_{1}^{-}$ state are shown in Fig. 1. 
The calculated energy levels are in good agreement 
with the experimental energy levels, except for $J^{\pi}=4^{+}_{2}$ and $J^{\pi}=6^{+}_{2}$ 
levels for $^{64}$Ge. 
In particular, the energy levels of $^{68}$Se are well reproduced. 
The experimental level sequence in the ground-state band
shows a quadrupole deformation. 

\begin{figure}[t]
\includegraphics[width=8cm,height=10cm]{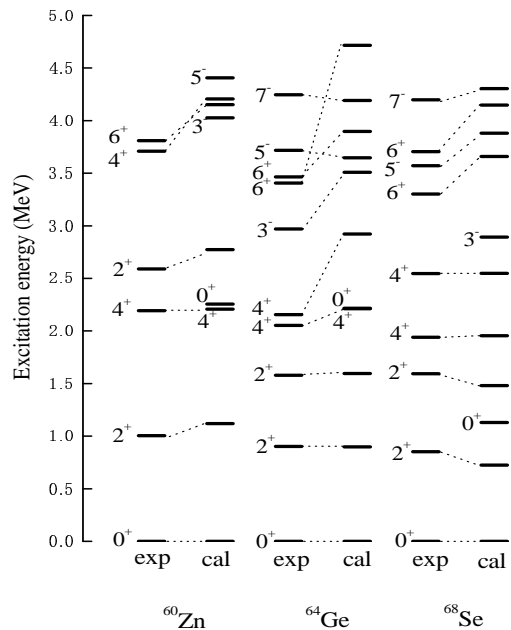}
  \caption{Experimental and calculated energy levels of $^{60}$Zn, $^{64}$Ge, and $^{68}$Se.}
  \label{fig1}
\end{figure}
\begin{figure}[t]
\includegraphics[width=8cm,height=10cm]{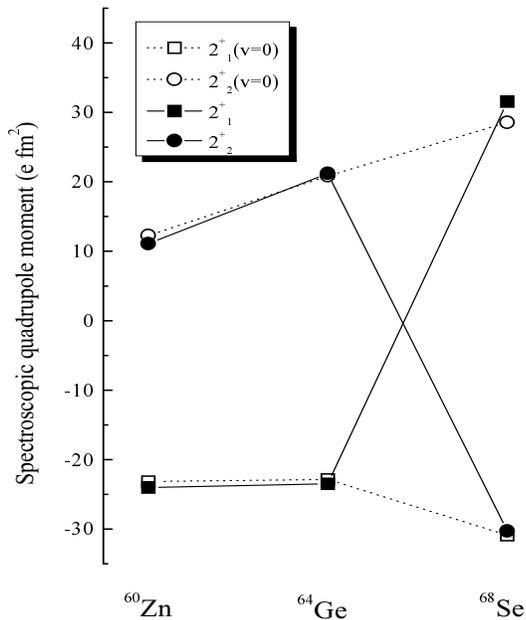}
  \caption{Spectroscopic quadrupole moments in $^{60}$Zn, $^{64}$Ge, and $^{68}$Se. 
  The solid squares and circles are quadrupole moments of the first and second excited 
  $2^{+}$ states, respectively, obtained by the shell model calculations. The open squares 
  and circles are those obtained by ignoring the $T=1$ monopole matrix elements. }
  \label{fig2}
\end{figure}

In order to study the quadrupole deformation of these nuclei, we examine the spectroscopic 
quadrupole moments and $E2$ transitions. 
We adopted the effective charge $e_{p}=1.5e$ for proton and $e_{n}=0.5e$ for neutron. 
Figure 2 shows the calculated spectroscopic quadrupole moments for the first and second 
excited $2^{+}$ states for $^{60}$Zn, $^{64}$Ge, and $^{68}$Se. 
The quadrupole deformation was estimated from the spectroscopic 
quadrupole moments for the first and second excited $2^{+}$ states on the assumption of 
axially symmetric deformation. 
For $^{60}$Zn and $^{64}$Ge, the deformation estimated from the quadrupole 
moments are prolate with $\beta\sim$ 0.2 for the first excited $2^{+}$  state ($2_{1}^{+}$)
and oblate with $\beta\sim - 0.2$ for the second excited $2^{+}$ state ($2_{2}^{+}$). 
The nucleus $^{68}$Se favors an oblate deformation  with $\beta\sim - 0.2$ since 
the value of intrinsic quadrupole moment for the $2_{1}^{+}$ state becomes negative. 
On the other hand, the $2_{2}^{+}$ state has prolate deformation with $\beta\sim$ 0.2. 
The results indicate shape transition in these nuclei. 
We examined the contributions of interaction matrix elements to the quadrupole moments. 
We found that the $T=1$ matrix elements between $f_{5/2}$ and 
$p_{1/2}$ orbits strongly affect the quadrupole moments. 
The $T=1$ monopole interaction is particularly important for the inversion of signs 
of the quadrupole moments in $^{68}$Se. As seen in Fig. (2), the inversion of the quadrupole 
moments occurs in $^{68}$Se, when we remove only the $T=1$ monopole matrix elements 
$v=V_{m}(f_{5/2},p_{1/2};T=1)= -0.4$ MeV from all of the interaction matrix elements. 
Namely, the shape of ground state in $^{68}$Se becomes prolate in this case. 
Thus, we can say from our calculations that the inverse signs of the quadrupole moments 
are attributed to the $T=1$ interaction with matrix element between $f_{5/2}$ and 
$p_{1/2}$ orbits. 

The quadrupole deformation of $^{64}$Ge estimated from the calculated 
$B(E2;2_{1}^{+}\rightarrow 0_{1}^{+})$ is $\beta\sim$ 0.2 \cite{Kaneko}. 
This value is in agreement with the predictions $\beta\sim$0.22 by M${\rm \ddot{o}}$ller 
and Nix \cite{Moller}, and $\beta\sim$ 0.22 by Ennis {\it et al.} \cite{Ennis}. 
The calculated value of $B(E2;2^{\dag}_{2}\rightarrow 
2^{\dag}_{1})/B(E2;2^{\dag}_{2}\rightarrow 0^{\dag}_{1})$ is very large, which indicates 
the $\gamma$-softness according to the Davydov model \cite{Davydov}. 
This is in agreement with the $\gamma$-softness or triaxiality 
estimated from the experimental data and the other theoretical models. 
It was demonstrated \cite{Kaneko} that the proton-neutron part $Q_{p}Q_{n}$ of the 
quadrupole-quadrupole interaction is important for the $\gamma$-softness or triaxiality. 
For $^{68}$Se, the quadrupole deformation estimated from the calculated 
$B(E2;2_{1}^{+}\rightarrow 0_{1}^{+})$ is $\beta\sim$ 0.21. 
The deformations estimated from the quadrupole moment and $B(E2)$ are smaller than 
$\beta\sim$ 0.27 estimated by Fischer {\it et al.} \cite{Fischer}, $\beta\sim$0.30 
obtained by Jenkins {\it et al.} \cite{Jenkins} for $^{69}$Se 
and the value $\beta\sim$ 0.33 calculated by Petrovici {\it et al.} \cite{Petrovici}. 
This can be attributed to neglection of excitations from the $f_{7/2}$ orbit 
because of computational limitation. 
\begin{figure}[b]
\includegraphics[width=9cm,height=10cm]{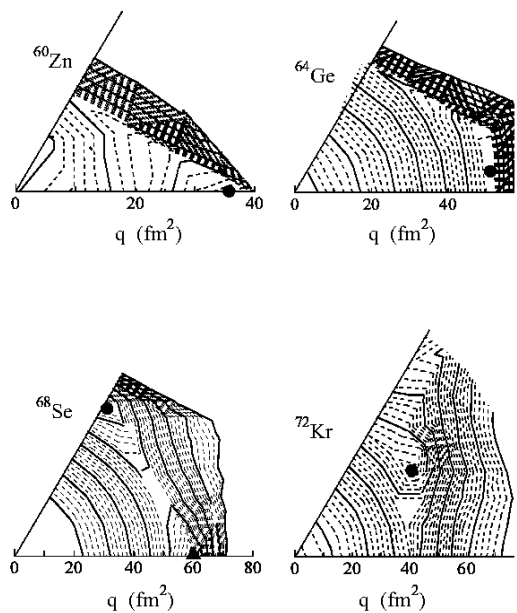}
  \caption{Contour plots of PES's on 
  $q$-$\gamma$ plane in the CHF calculations for $^{60}$Zn, $^{64}$Ge, $^{68}$Se, 
  and $^{72}$Kr. The solid circles denote the PES minima, and the solid triangle 
  the second minimum.}
  \label{fig3}
\end{figure}

\begin{figure}[b]
\includegraphics[width=8cm,height=10cm]{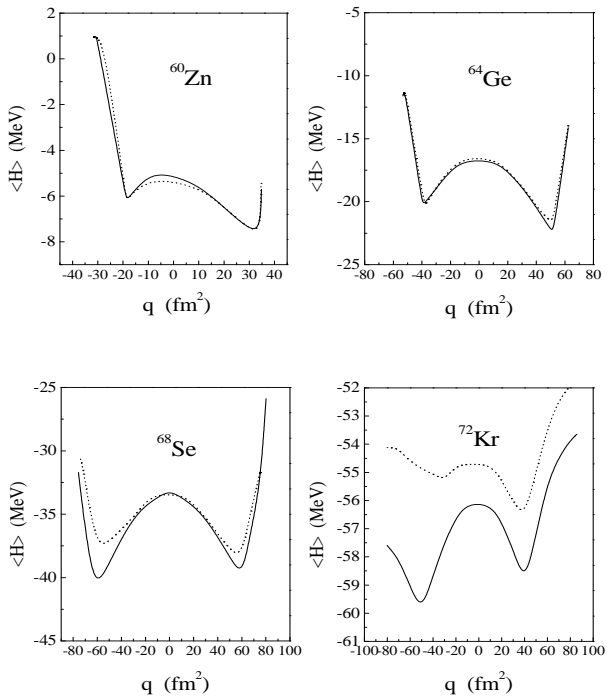}
  \caption{PES's as a function of $q$ along the axially symmetric line. 
  The solid curves are the PES's in the full shell model 
  matrix elements and the dotted curves are those obtained by ignoring the $T=1$ 
  monopole matrix element between $f_{5/2}$ and $p_{3/2}$.}
  \label{fig4}
\end{figure}

Let us next examine the nuclear shapes including triaxiality of $^{60}$Zn, 
$^{64}$Ge, $^{68}$Se, and $^{72}$Kr by an alternative approach, the CHF method \cite{Mizusaki}, 
which is carried out by the following quadratic way: 
\begin{eqnarray}
 H' & = & H + \alpha\sum_{\mu}(\langle Q_{2\mu}\rangle - q_{\mu})^{2}
            + \beta(\langle J_{x}\rangle - j_{x})^{2}, 
\end{eqnarray}
where $J_{x}$ are the $x$-component of the angular momentum operator. 
The $q_{\mu}$'s are the constant parameters 
$q_{0}=\sqrt{\frac{5}{16\pi}}q{\rm cos}\gamma$, 
$q_{\pm 2}=\sqrt{\frac{5}{16\pi}}q{\rm sin}\gamma$, and $q_{\pm 1}=0$, where 
$q$ is the isoscalar intrinsic quadrupole moment and $\gamma$ is the triaxial angle. 
We set $j_{x} = \sqrt{J(J+1)}$ with $J$ being the total angular momentum of the 
state. 
The parameters, $\alpha$ and $\beta$, are taken so as to achieve 
convergence of the iteration of the gradient method. 
Then, potential energy surface (PES) is defined as the expectation value 
$\langle H \rangle$ with respect to the CHF state for a given $q$ and $\gamma$. 
Figure 3 shows the contour plots of the PES in the $q$-$\gamma$ plane for $^{60}$Zn, 
$^{64}$Ge, $^{68}$Se, and $^{72}$Kr. 
We can see remarkable shape changes of these nuclei. 
The PES minimum exhibits prolate deformation for $^{60}$Zn and $^{64}$Ge, and 
oblate one for $^{68}$Se, and triaxial one for $^{72}$Kr. 
The characteristic features of the PES's are the $\gamma$-softness for $^{64}$Ge and 
oblate-prolate coexistence for $^{68}$Se, which are consistent with the previous 
discussions \cite{Ennis,Fischer}. 
Moreover, we calculated an angular-momentum projected PES which is 
obtained by calculating the expectation value $\langle H \rangle$ with respect to the 
angular-momentum projected CHF wave function for a given $q$ and $\gamma$. 
The angular-momentum projection method improves considerably the ground-state energies. 
The angular-momentum projected PES's show a similar potential energy surface 
to the CHF. 

The PES's as a function of $q$ along the axially symmetric line are shown for 
$^{60}$Zn, $^{64}$Ge, $^{68}$Se, and $^{72}$Kr in Fig. 4. This figure shows the aspect 
of shape coexistence and shape changes in these nuclei, from different angle. 
Figure 4 also shows that if the $T=1$ monopole interactions are removed, the minima 
of the PES's in $^{68}$Se and $^{72}$Kr shift to the prolate side, while it remains 
unchanged in $^{60}$Zn, $^{64}$Ge. Thus the $T=1$ monopole 
interactions with the matrix element $V_{m}(f_{5/2},p_{1/2};T=1)$ are confirmed to be the 
oblate-driving force for $^{68}$Se and $^{72}$Kr. 
The shape of the ground state becomes oblate for $^{68}$Se
 where the Fermi energy for 
neutron lies between the $f_{5/2}$ and $p_{1/2}$ orbits. 
 The CHF calculation shows that the 
shape of $^{72}$Kr is triaxial, while it is oblate in the HFB result
\cite{Yamagami} 
and the TRS results \cite{Angelis}. Note that the PES of $^{72}$Kr
is flat toward oblate 
shape and shows gamma softness, as seen in Fig. 3. 

\begin{figure}[t]
\includegraphics[width=8cm,height=10cm]{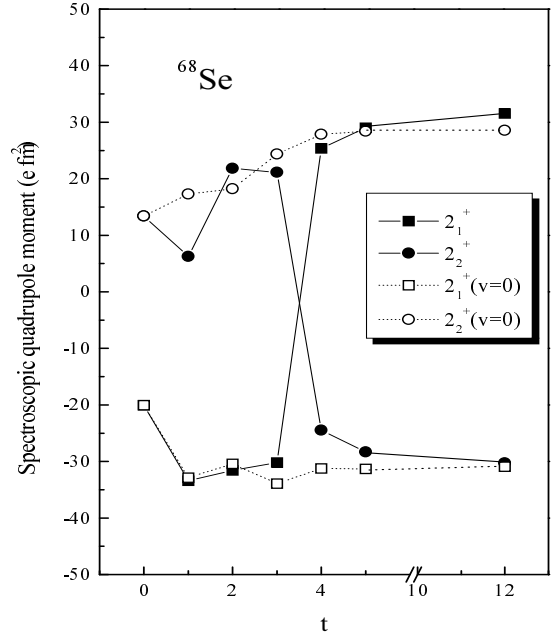}
  \caption{Spectroscopic quadrupole moment as a function of $t$. 
           The solid squares and circles are  
           for the first and second excited $2^{+}$ states, 
           respectively. The open squares and circles are those obtained by ignoring the 
           $T=1$ monopole matrix elements.}
  \label{fig5}
\end{figure}

We next analyze how excitations from the orbits $(p_{3/2},f_{5/2})$ to $(p_{1/2},g_{9/2})$ 
contribute to the oblate deformation. 
We calculate the quadrupole moments by enlarging configuration space in the truncation 
scheme $\bigoplus_{s\leq t}(p_{3/2}f_{5/2})^{A-56-s}(p_{1/2}g_{9/2})^{s}$, 
where $t$ is the maximum number of particles allowed to be excited. 
Figure 5 shows the spectroscopic quadrupole moments as a function of $t$ 
for $^{68}$Se. 
The spectroscopic quadrupole moment is negative for $2_{1}^{+}$ for $2_{2}^{+}$ 
when $t \le 3$, but suddenly their signs change at $t$=4. 
This means that the four-particle four-hole ($4p-4h$) excitations from 
$(p_{3/2}f_{5/2})$ to $(p_{1/2}g_{9/2})$ are very important for the oblate configurations. 
Such a change does not occur when the $T=1$ monopole interactions are removed, 
as shown by the dotted lines in Fig. 4. 
Table I shows the occupation numbers of the $2p_{3/2}$, $1f_{5/2}$, $2p_{1/2}$, 
and $1g_{9/2}$ orbits for the first and second excited $2^{+}$ states of $^{60}$Zn 
and $^{68}$Se in the cases of the $T=1$ monopole force strengths $v=-0.4$ MeV and 0.0 MeV. 
When we add the $T=1$ monopole force, the occupation number of $p_{1/2}$ for the first 
excited $2^{+}$ state for $^{68}$Se increases remarkably. 
Thus the $T=1$ monopole force contributes to $4p-4h$ excitations from $(p_{3/2}f_{5/2})$ to 
$p_{1/2}$, which is especially important for the oblate configuration. 

  Our conclusion is different from that of the VAMPIR calculations \cite{Petrovici},
where the oblate-driving force for $^{68}$Se is in monopole shifts for $T=0$ 
matrix elements between the $g_{9/2}$ and $f_{5/2}(f_{7/2})$ orbits.
We examined this question by changing monopole shifts and
single-particle energy $\varepsilon_{g9/2}$.
The investigations of the $Q$ moments show that the monopole shift $V_m(f_{5/2},g_{9/2};T=0)$
hardly contributes to the oblate deformation under the single-particle 
energies adopted in this paper. 
If we lower $\varepsilon_{g9/2}$ below $\varepsilon_{p1/2}$ as done
in Ref. \cite{Petrovici}, many nucleons occupy the $g_{9/2}$ orbit
and $V_m(f_{5/2},g_{9/2};T=0)$ contributes to the oblate deformation.
 The nucleon occupation numbers of the $p_{1/2}$ and $g_{9/2}$
orbits and phases of respective components of wave-functions affects
the sign of the $Q$ moments.
It seems that the VAMPIR conclusion depends on the inversion of $\varepsilon_{p1/2}$
and $\varepsilon_{g9/2}$.

\begin{table}[t]
\caption{Occupation numbers in the first and second excited $2^{+}$ states 
         for $^{60}$Zn and $^{68}$Se.}
\begin{tabular}{c|cccc|cccc} 
            & \multicolumn{4}{c}{$2_{1}^{+}$} & \multicolumn{4}{|c}{$2_{2}^{+}$} \\
 $^{60}$Zn  & $p_{3/2}$  &  $f_{5/2}$ & $p_{1/2}$  & $g_{9/2}$ &  $p_{3/2}$  &  $f_{5/2}$ & $p_{1/2}$  & $g_{9/2}$ \\  \hline
 $v$=-0.4   & 0.98     & 0.51     & 0.49     & 0.02    &  1.08     &  0.59    & 0.24     & 0.09      \\
 $v$= 0.0   & 1.06     & 0.46     & 0.46     & 0.02    &  1.11     &  0.56    & 0.24     & 0.09      \\
   \\
 $^{68}$Se  & $p_{3/2}$  &  $f_{5/2}$ & $p_{1/2}$  & $g_{9/2}$ &  $p_{3/2}$  &  $f_{5/2}$ & $p_{1/2}$  & $g_{9/2}$ \\  \hline
 $v$=-0.4   & 2.31     & 2.38     & 1.20     & 0.11    &  2.37     &  2.81    & 0.69     & 0.13      \\
 $v$= 0.0   & 2.65     & 2.78     & 0.42     & 0.15    &  2.70     &  2.33    & 0.80     & 0.15      \\
\end{tabular}
\label{table2}
\end{table}

Finally, we discuss shortly standard $P+QQ$ force model calculations neglecting the octupole 
force and the monopole shifts. We set the force parameters such as 
so as to reproduce the energy levels of ground-state band for 
$^{68}$Se. Note that the quadrupole force strength is larger than that in Eq. (4). 
Then, we obtain the positive spectroscopic quadrupole moment for the $2_{1}^{+}$ state and 
negative one for the $2_{2}^{+}$ state. 
Similarly in the CHF calculations, the first PES minimum is in the oblate side and the 
second minimum is in the prolate side. 
However, the standard $P+QQ$ force model cannot consistently reproduce the energy levels of 
the side band for $^{68}$Se and the ground-state bands for $^{60}$Zn and $^{64}$Ge. 

In conclusion, we investigated nuclear shape transition and oblate-prolate coexistence 
in $N=Z$ $fpg$-shell nuclei. We analyzed the quadrupole moments in the shell model and 
the CHF calculations. The analysis shows that the $T=1$ monopole interactions with the 
matrix element between $f_{5/2}$ and $p_{1/2}$ orbits play an important role in producing 
the oblate shape for $^{68}$Se and $^{72}$Kr, while they do not play the same role for 
$^{60}$Zn and $^{64}$Ge. 
Note that this conclusion is different from that of the VAMPIR calculations by Petrovici 
{\it et al.}, where the origin of oblate-driving force is $T=0$ monopole matrix elements 
between $g_{9/2}$ and $f_{5/2}(f_{7/2})$ orbits. However, in our calculation we could not 
find any large influence from these $T=0$ monopole matrix elements on the oblate deformation. 
The shape transition and oblate-prolate coexistence in the CHF calculations 
are consistent with those of the Skyrme HFB calculations \cite{Yamagami}. The present 
calculations show that the oblate deformation originates in the $4p-4h$ excitations 
from $(p_{3/2}f_{5/2})$ to $(p_{1/2}g_{9/2})$. We can expect that 
the $T=1$ monopole interactions between $f_{5/2}$ and $p_{1/2}$ orbits are also important 
for the oblate or triaxial deformation in $N\neq Z$ nuclei where the Fermi surface for neutron 
lies between $f_{5/2}$ and $p_{1/2}$ in $fpg$ shell. Further investigations are in progress.



\end{document}